# Imagined-Trailing-Whitespace-Agnostic Levenshtein Distance For Plaintext Table Detection

**Kartik Vempala**                                                   `kartikvempala@gmail.com`
*Bloomberg LP*
*New York, NY, USA*

**Abstract**

The standard algorithm for Levenshtein distance, treats trailing whitespace the same as any other letter or symbol. However, when humans compare 2 strings, we implicitly assume that both strings are padded by infinite trailing whitespace. This informs our expectations for what the costs for insertion, deletion and replacement, should be. This violation of our expectations results in non-intuitive edit distance values. To account for this specific human intuition, a naive approach which considers "all possible" substrings of trailing whitespace would yield an $O(n^3)$ algorithm. In this work, we provide an efficient $O(n^2)$ algorithm to compute the same.

## 1 Introduction

A common approach to quantify the distance between 2 character strings, is to use the Levenshtein Distance (aka Edit Distance) (Levenshtein, 1966) algorithm and it's many variants. Many of these variants have been motivated by wanting to close the gap between intuitive expectations of the edit distance, versus the actual distances obtained by the original formulation of the algorithm.

Broadly, these variants fall under a few categories (Hládek et al., 2020; Varol et al., 2010):
- Phonemic similarity (Kukich, 1992; Veronis, 1988)
- Keyboard layout (Brill & Moore, 2000)
- Domain-specific word frequencies (Hauser & Schulz, 2007; Tillenius, 1996)

To our knowledge, one commonly occurring category that has not enjoyed a lot of attention from researchers, is the implicit assumption by humans, to imagine infinite whitespace padded at the ends of strings being compared.

## 2 Motivation

Tables are frequently used in documents to communicate multi-dimensional information in a way that is compact and easy to understand. They also provide a convenient format to programmatically extract the information therein. When used in structured document types like HTML and Rich-Text, the tables are relatively easy to detect and parse. However, this can be extremely challenging to do for plain-text documents.

Levenshtein distances are frequently used to detect the boundaries of tables in a document (Hu et al., 2002; Zanibbi et al., 2004). This works well when the rows of the table have roughly the same length and when the rightmost cells in rows are not missing or empty. However, when the rightmost cells are missing, the distance values we get, are unintuitive. Consider the following plain-text table, where the 2nd row is missing "weight" information:

**Original Text**

```
Bill Nye    6 ft 0 inches    190 lb
Tina Fey    5 ft 5 inches
Mike Fox    5 ft 4 inches    130 lb
```

**Normalized Text (obtained by replacing letters with 'a', and numbers with '9')**

```
aaaa aaa    9 aa 9 aaaaaa    999 aa
aaaa aaa    9 aa 9 aaaaaa
aaaa aaa    9 aa 9 aaaaaa    999 aa
```

Consider the first 2 rows of the normalized text. Assuming a cost of 1 for insertion, deletion, and replacement, the Levenshtein distance between row 1 and row 2, is 9.

However, intuitively, we only expect a distance of 5. A distance of 3 for substituting "999" with the imagined trailing whitespace in the 2nd row; plus a distance of 2 for substituting "aa".

This intuition can be formalized by observing that when humans compare the rows of the table, we imagine the rows to be padded with infinite trailing whitespace. In this work, we provide an efficient $O(n^2)$ algorithm to calculate the Levenshtein distance based on this formalization.

For further illustration, consider the examples in the table below:

| String 1 | String 2 | Actual Distance | Expected Distance |
|---|---|---|---|
| 'A B C' | 'A B C **D** ' | 3 | 1 |
| 'A B C' | 'A B C             **D**' | 14 | 1 |
| 'A B C    ' | 'A B C          **D**' | 11 | 1 |
| 'A B C' | 'A B C      **x**     **D**' | 11 | 2 |
| 'A B C' | 'A B **D**       ' | 5 | 1 |
| 'A B C' | 'A B **D**   **E**    **F**      ' | 12 | 3 |

More elaborate examples, with various costs for insertion/deletion and replacement, can be found in **Appendix A**.

## 3 Algorithm

In this section, we first describe, in abstract, the key idea behind the algorithm. This is followed by an $O(n^2)$ Python-like pseudocode implementation of the same.

### 3.1 Key idea and Explanation

The key idea is to append a symbol denoting "infinite whitespace" (denoted by '∞') to the end of both strings.

The edit distance algorithm proceeds as usual, until '∞' is encountered in one or both strings. If the '∞' symbol of one string is being compared with the '∞' of the other string, they match with a cost of 0, and the algorithm terminates.

Let's assume, without loss of generality, that the '∞' of the first string, is being compared with a non-'∞' letter of the second string. If '∞' is being compared with whitespace, they match with a cost of 0, and the algorithm continues without incrementing the index of the first string. If '∞' is being compared with any other letter (say, 'x'), then the insertion/deletion and replacement costs are the same as though 'x' were being compared with whitespace. The algorithm continues without incrementing the index of the first string

More formally, we define it as the following:

$$lev(a,b) = \begin{cases} lev(tail(a), tail(b)) & \text{if } a[0] = b[0] \\ lev(tail(a), b) & \text{if } a[0] = \text{'\,'} \text{ and } b[0] = \infty \\ lev(a, tail(b)) & \text{if } a[0] = \infty \text{ and } b[0] = \text{'\,'} \\ 1 + \min \begin{cases} lev(a, tail(b)) \\ lev(tail(a), b) \\ lev(tail(a), tail(b)) \end{cases} & \text{otherwise.} \end{cases}$$

where

$$tail(s) = \begin{cases} \{\infty\} & \text{if } length(s) \leq 1 \\ substring(s) & \text{if } length(s) > 1 \end{cases}$$

where

- substring(s) of a string s, is a string with all but the 1st letter of s
- {a, b, c, ...} is a sequence of letters in a string

### 3.2 Pseudocode

The following is an efficient $O(n^2)$ implementation of the algorithm in Python-like pseudocode.

Note that we don't actually have to append an '∞' symbol to the strings. The indices of the string are sufficient.

```
def editDistance(string s1, string s2):
  length1 = len(s1)
  length2 = len(s2)
  d = <matrix of shape (length1+1) x (length2+1)>
  # initialize cost matrix
  for i = 1 ... length1:
    d[i][0] = i
    for j = 1 ... length2:
      d[0][j] = j
  d[0][0]=0
  for i = 1 ... length1:
    for j = 1 ... length2:
      insert1Cost = d[i-1][j] + costOfInserting(s1[i])
      insert2Cost = d[i][j-1] + costOfInserting(s2[j])
      replaceCost = costOfReplacing(s1[i], s2[j])
      if j == lenstr2: # reached '∞'
```

```
        insert1Cost = min(insert1Cost,
                          d[i-1][j]
                          + costOfReplacing(s1[i-1], ' '))
      if i == lenstr1: # reached '∞'
        insert2Cost = min(insert2Cost,
                          d[i][j-1]
                          + costOfReplacing(s2[j-1], ' '))
      d[i][j] = min(insert1Cost, insert2Cost, replaceCost)
  return d[length1][length2]
```

## 4 Conclusion and Future Work

In this work, we add to the existing body of research for adapting the vanilla Levenshtein distance to be more in line with human intuition and expectation. We consider the as-yet-unexplored area of accounting for imagined trailing whitespace, and describe an efficient $O(n^2)$ algorithm to compute the same.

A potential avenue for future research might be to extend the current formulation to account for infinite leading whitespace as well, which might be helpful for string alignment use-cases.

## 5 Appendix A

In the examples below, lowercase letters are normalized to "a", uppercase letters are normalized to "A", digits are normalized to "9", and special symbols are kept as-is.

### 5.1 Replacement Costs

We assume the following costs for replacing one letter with another.

|   | a | A | 9 | ( | ) | , | $ | \<space\> |
|---|---|---|---|---|---|---|---|---|
| **a** | 0 | 2 | 4 | 999 | 999 | 999 | 999 | 1 |
| **A** | 2 | 0 | 4 | 999 | 999 | 999 | 999 | 1 |
| **9** | 4 | 4 | 0 | 999 | 999 | 999 | 999 | 4 |
| **(** | 999 | 999 | 999 | 0 | 999 | 999 | 999 | 999 |
| **)** | 999 | 999 | 999 | 999 | 0 | 999 | 999 | 999 |
| **,** | 999 | 999 | 999 | 999 | 999 | 0 | 999 | 999 |
| **$** | 999 | 999 | 999 | 999 | 999 | 999 | 0 | 999 |
| **\<space\>** | 1 | 1 | 4 | 999 | 999 | 999 | 999 | 0 |

### 5.2 Insertion/Deletion Costs

We assume a cost of 1, for insertion/deletion of all letters.

#### 5.2.1 Example 1

**Original**
```
nameA   30   62
nameB
nameC        73
```

**Normalized**

```
aaaaA   99  99
aaaaA
aaaaA       99
```

**Edit Distances**

|                | Actual Distance | Expected Distance |
|----------------|-----------------|-------------------|
| Row 2 vs Row 1 | 8               | 4                 |
| Row 2 vs Row 3 | 8               | 2                 |

### 5.2.2  Example 2

**Original**

```
Value1  No   6   01 02     True
Value2  Yes  7             False
Value3  No   6   01 01
Value4  No   6
```

**Normalized**

```
Aaaaa9  Aa   9   99 99    Aaaa
Aaaaa9  Aaa  9            Aaaaa
Aaaaa9  Aa   9   99 99
Aaaaa9  Aa   9
```

**Edit Distances**

|                | Actual Distance | Expected Distance |
|----------------|-----------------|-------------------|
| Row 3 vs Row 1 | 8               | 4                 |
| Row 3 vs Row 2 | 17              | 10                |
| Row 3 vs Row 4 | 8               | 4                 |
| Row 4 vs Row 1 | 16              | 8                 |
| Row 4 vs Row 2 | 18              | 6                 |

### 5.2.3  Example 3

**Original**

```
PPHH 123     Principles of physics    (2)      Fall 2005
ENGL 456H    Critical thinking        (6)
ENGL 100     Essay composition        (7)
```

**Normalized**

```
AAAA 999     Aaaaaaaaaa aa aaaaaaa    (9)      Aaaa 9999
AAAA 999A    Aaaaaaaa aaaaaaaa        (9)
AAAA 999     Aaaaa aaaaaaaaaaa        (9)
```

**Edit Distances**

|                | Actual Distance | Expected Distance |
|----------------|-----------------|-------------------|
| Row 2 vs Row 1 | 19              | 16                |
| Row 3 vs Row 1 | 19              | 15                |

### 5.2.4 Example 4

**Original**

```
Gamestop    111111    4, 3, 14, 16, 17, 18       654,321
Amc         222222    4, 14
Tesla       333333    4, 11, 14               98,760,000
```

**Normalized**

```
Aaaaaaaa    999999    9, 9, 99, 99, 99, 99       999,999
Aaa         999999    9, 99
Aaaaa       999999    9, 99, 99               99,999,999
```

**Edit Distances**

|              | Actual Distance | Expected Distance |
|---|---|---|
| Row 2 vs Row 1 | 29 | 24 |
| Row 2 vs Row 3 | 29 | 15 |

### 5.2.5 Example 5

**Original**

```
Gamestonk banking   2021    234,012.50     $11,121
Short sellers inc   2020    345,280.00
Dogecoin partners   2023                   $12,314
```

**Normalized**

```
Aaaaaaaaa aaaaaaa   9999    999,999.99     $99,999
Aaaaa aaaaaaa aaa   9999    999,999.99
Aaaaaaaa aaaaaaaa   9999                   $99,999
```

**Edit Distances**

|              | Actual Distance | Expected Distance |
|---|---|---|
| Row 2 vs Row 1 | 15 | 10 |
| Row 2 vs Row 3 | 23 | 20 |